\documentclass[epj]{svjour}
\usepackage{epsf,epsfig,float,amssymb,latexsym}
\usepackage{graphics,psfrag}
\newcommand{\ck}{{\cal K}}

\newcommand{\Op}{{\mathcal{O}}(p)}
\newcommand{\Opd}{{\mathcal{O}}(p^2)}

\newcommand{\Opc}{{\mathcal{O}}(p^4)}

\newcommand{\be}{\begin{equation}}
\newcommand{\ee}{\end{equation}}
\newcommand{\ba}{\begin{eqnarray}}
\newcommand{\ea}{\end{eqnarray}}

\newcommand{\nn}{\nonumber}

\newcommand{\vs}{\vspace{-0.0cm}}
\newcommand{\g}{\gamma}
\begin{document}
\title{\hfill{{\small CAFPE 79/06 , UGFT-209/06}}\\
 Aspects of Strangeness $-$1 Meson-Baryon Scattering}
\author{Jos\'e A. Oller\inst{1}, 
Joaquim Prades\inst{2} \and Michela Verbeni\inst{1} % etc
}                     % Do not remove
\institute{Departamento de F\'{\i}sica, Universidad de
Murcia, E-30071 Murcia, Spain 
 \and CAFPE and Departamento de F\'\i sica Te\'orica y del Cosmos,
Universidad de Granada,\\
Campus de Fuente Nueva, E-18002 Granada, Spain}
\date{Received: date / Revised version: date}
\abstract{We consider meson-baryon interactions in S-wave with strangeness $-1$. 
This is a sector
populated by plenty of resonances interacting in several two-body coupled channels.  We
 consider a large set of experimental data, where the recent experiments are
 remarkably accurate. This requires a
sound theoretical description to account for all the data and we employ Unitary Chiral
Perturbation Theory up to and including ${\Opd}$. 
 The spectroscopy of our solutions is studied  within this approach, discussing  the rise 
  of the two $\Lambda(1405)$ resonances and of 
 the  $\Lambda(1670)$, $\Lambda(1800)$, $\Sigma(1480)$, $\Sigma(1620)$ and
 $\Sigma(1750)$. We finally argue about our preferred fit.
\PACS{
      {13.75.Jz}{Kaon-baryon interactions}   \and
      {12.39.Fe}{Chiral Lagrangians}  \and
      {11.80.-m}{Relativistic scattering theory}   \and
      {11.80.Gw}{Multichannel scattering}}% end of PACS codes
      } %end of abstract
\maketitle
\section{Introduction}
\label{intro}

The study of strangeness $-1$ meson-baryon dynamics comprising the $\bar{K}N$ plus
coupled channels, has been renewed both from theoretical and experimental sides.
Experimentally, we have new exciting data like the increasing improvement in 
precision of the measurement of the $\alpha$ line of kaonic hydrogen accomplished recently
by DEAR \cite{DEAR}, and its foreseen better determination, with an expected error of a few eV,
by the DEAR/SIDDHARTA Collaboration \cite{sid}. This has established a
challenge to theory in order to match such precision. In this respect, ref.\cite{akaki} provides 
an improvement over the traditional Deser formula for relating scattering at threshold
with the spectroscopy of hadronic atoms \cite{deser}.
 In addition, one needs a good scattering
amplitude to be implemented in this equation. The study of strangeness $-1$ meson-baryon 
 scattering has a long history 
\cite{dalitz,galileo,martin,juelich,hamaie,landau,cloudy,schat} within K-matrix models,
 dispersion relations, meson-exchange
models, quark models, cloudy bag-models or large $N_c$ QCD, just to quote a few.
 However, in more recent years it has 
received a lot of attention from the application of SU(3) baryon Chiral Perturbation Theory (CHPT) to this
sector together with a unitarization procedure, see e.g., 
\cite{kaisersiegel,npa,oset,reportramos,ollerm,teamL,lutznieves,bura,reply,opv}. Recently, 
ref.\cite{akaki} pointed out the possible inconsistency of the DEAR measurement on kaonic hydrogen
 and $K^-p$ scattering,
since the unitarized CHPT results, able to reproduce the scattering data, 
were not in agreement with DEAR. Later on, ref.\cite{bura} insisted on
this fact based on its own fits, although they only included partially 
the ${\Opd}$ CHPT amplitudes \cite{reply}.
 However, the situation
changed with ref.\cite{opv}, as it was shown that one can obtain fits in Unitary CHPT (UCHPT),  
including full ${\Opd}$ CHPT amplitudes, which are compatible both with DEAR and with $K^-p$ scattering
data. In addition, ref.\cite{oepja} extended the work of ref.\cite{opv} 
by including additional experimental data, recently
measured with remarkable precision by the Crystal Ball Collaboration, for the reactions 
$K^-p\rightarrow \eta \Lambda$ \cite{nefkens} and  $\pi^0\pi^0\Sigma^0$ \cite{prakhov}. 
The importance of
including the latter data in any analysis of $K^-p$ interactions has been singled out 
in ref.\cite{magas}. We will report here about the series of works \cite{opv,reply,oepja} 
on meson-baryon CHPT. Recently, we also presented the first full and minimal SU(3) CHPT 
meson-baryon Lagrangians to ${\cal O}(p^3)$ in ref.\cite{lag}.

The study of $K^-p$ plus coupled channel interactions offers, from the theoretical point of view, a very
challenging test ground for chiral effective field theories of QCD since one has there plenty of
experimental data, Goldstone bosons dynamics and large and explicit SU(3) breaking. In addition, this 
sector shows a very rich spectroscopy with many I=0, 1 S-wave resonances that will be the object 
of our study as well. Apart from that,
these interactions are interrelated with many other interesting areas, 
e.g., possible kaon condensation in neutron-proton stars,
 large yields of $K^-$ in heavy ions collisions, kaonic atoms
  or non-zero strangeness content of the proton. 

\begin{figure}
\psfrag{a}{\begin{tabular}{l} {\small ${\Op}$}\\ {\small Seagull} \end{tabular}}
\psfrag{b}{\begin{tabular}{l} {\small ${\Op}$}\\ {\small Direct } \end{tabular}}
\psfrag{c}{\begin{tabular}{l} {\small ${\Op}$}\\ {\small Crossed } \end{tabular}}
\psfrag{d}{\begin{tabular}{l} {\small ${\Opd}$}\\ {\small Contact terms} \end{tabular}}
\centerline{\epsfig{file=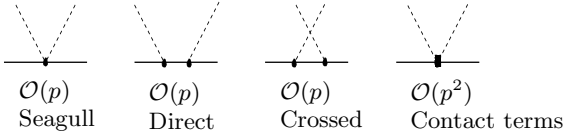,height=.6in,width=2.5in,angle=0}}
\vspace{0.2cm}
\caption[pilf]{\protect \small
 Diagrams for the calculation of the baryon CHPT scattering amplitudes up to 
 and including ${\Opd}$. The first three diagrams are of ${\Op}$ while the last one is 
 of ${\Opd}$. 
\label{fig1}}
\end{figure}
  
\section{Formalism and Results}
\label{sec:1}
A general meson-baryon partial wave in coupled channels can be written in matrix 
notation as \cite{ollerm},
\be
T=\left[1+{\cal K}g\right]^{-1}{\cal K}~,
\label{eqt}
\ee
where $g$ is a diagonal matrix that comprises the unitarity bubble for every channel and 
${\cal K}$ is the interaction kernel that is determined from meson-baryon CHPT. This is accomplished 
by performing a power expansion of $T$ calculated from CHPT and then matched, order by order,
 with the chiral expansion of eq.(\ref{eqt}),
\be
T_1+T_2+T_3+\Opc={\cal K}_1+\ck_2+\ck_3-\ck_1 \cdot g \cdot \ck_1+\Opc~,
\label{texp}
\ee
taking into account that $g$ is of chiral order one. We calculate $\ck$ up to ${\Opd}$, 
$\ck_1=T_1$ and $\ck_2=T_2$. The lowest order result, $T_1$, contains the seagull, direct and crossed
exchange diagrams, while the next-to-leading order amplitudes, $T_2$, come from pure contact interactions. 
This is shown diagramatically in fig.\ref{fig1}. Once the kernel $\ck=T_1+T_2$ has been calculated,
 we insert it in eq.(\ref{eqt}) and evaluate the S-wave amplitudes.

In ref.\cite{opv} a large set of meson-baryon scattering data was fitted which 
was enlarged   in
ref.\cite{oepja} including new precise ones from the Crystal Ball Collaboration. Namely, ref.\cite{opv} 
took into account  the 
$\sigma(K^- p\rightarrow K^- p)$ elastic cross section \cite{26plb,27plb,28plb,31plb}, 
the $\sigma(K^- p\rightarrow \bar{K}^0n)$ charge exchange one
\cite{26plb,27plb,31plb,29plb,30plb}, 
 and several hyperon production reactions, 
$\sigma(K^- p\rightarrow \pi^+\Sigma^-)$ \cite{26plb,27plb,28plb}, 
$\sigma(K^- p\rightarrow \pi^-\Sigma^+)$ \cite{27plb,28plb,31plb},  
$\sigma(K^- p\rightarrow \pi^0\Sigma^0)$ \cite{27plb} 
and  $\sigma(K^- p\rightarrow \pi^0\Lambda)$ \cite{27plb}.  In our normalization the corresponding 
cross section, keeping only the S-wave, is given by
\be
\sigma(K^-p \to M B)=\frac{1}{16 \pi s}\,\frac{p'}{p}\,|T_{K^- p\to MB}|^2~,
\label{KPMB}
\ee
where $MB$ denotes the final meson-baryon system, $p'$ the final CM three-momentum and 
$p$ the initial one. 

\begin{table}[H]
\begin{center}
\caption{A-type fits that agree with the DEAR data, eq.(\ref{deardata}). The  
$\sigma_{\pi N}$ value enforced in the fits is given in the 
first row. For precise definitions of the parameters $f$, $b_0$, $b_D$, $b_F$, $b_i$ and $a_i$ 
 see ref.\cite{oepja}. Three among the parameters
 $b_0$, $b_D$, $b_F$ and $b_i$ are fixed.
 \label{table:a4pnewvalues}}
\begin{tabular}{|l|l|r|r|r|}
\hline
Units& $\sigma_{\pi N}$ & $20^*$ & $30^*$ & $40^*$  \\
     & MeV           &     &      &   \\
\hline
MeV &         $f$      & $75.2$    & $71.8$    & $67.8$   \\
GeV$^{-1}$ & $b_0$     & $-0.615$  & $-0.750$  & $-0.884$    \\
GeV$^{-1}$ & $b_D$     & $+0.818$  & $+0.848$  & $+0.873$   \\
GeV$^{-1}$ & $b_F$     & $-0.114$  & $-0.130$  & $-0.138$        \\
GeV$^{-1}$ & $b_1$     & $+0.660$  & $+0.670$   & $+0.676$   \\
GeV$^{-1}$ & $b_2$     & $+1.144$  & $+1.169$   & $+1.189$   \\
GeV$^{-1}$ & $b_3$     & $-0.297$  & $-0.316$  & $-0.315$    \\
GeV$^{-1}$ & $b_4$     & $-1.048$  & $-1.181$  & $-1.307$  \\
	  & $a_1$      & $-1.786$  & $-1.591$  & $-1.413$  \\
	  & $a_2$      & $-0.519$  & $-0.454$  & $-0.386$  \\
	  & $a_5$      & $-1.185$  & $-1.170$  & $-1.156$  \\
	  & $a_7$      & $-5.251$  & $-5.209$  & $-5.123$  \\
	  & $a_8$      & $-1.316$  & $-1.310$  & $-1.308$  \\
	  & $a_9$      & $-1.186$  & $-1.132$  & $-1.050$  \\
 \hline
\end{tabular}
\end{center}
\end{table}

In addition, we also fit the precisely measured ratios 
at the $K^- p$ threshold \cite{nowak,tovee}:
\ba
\gamma&=&\frac{\sigma(K^-p\rightarrow \pi^+\Sigma^-)}
{\sigma(K^-p\rightarrow \pi^-\Sigma^+)}=2.36\pm 0.04~,\\
R_c&=&\frac{\sigma(K^-p\rightarrow \hbox{charged particles})}
{\sigma(K^-p\rightarrow \hbox{all})}=0.664\pm0.011~,\nn\\
R_n&=&\frac{\sigma(K^-p\rightarrow \pi^0\Lambda)}
{\sigma(K^-p\rightarrow \hbox{all neutral states})}=0.189\pm 0.015.
\nn
\label{ratios}
\ea
  The first two ratios,
which are Coulomb corrected, are measured with 1.7$\%$ precision, 
 which is of the same order as  the expected isospin violations. 
Indeed, all the other observables  we fit have
uncertainties larger than $5\%$. 

 Since we are just considering the S-wave amplitudes,  we  only include in the
fits those data points for the several $K^-p$ cross  sections 
with  laboratory frame   $K^-$ three-momentum  $p_K\leq 0.2$ GeV.
 This also enhances 
the sensitivity  to the lowest energy region in which  we are particularly 
interested.  We also 
 include in the fits the $\pi^{\pm}\Sigma^{\mp}$ 
event distributions from the chain of reactions $K^-p \to \Sigma^+(1660) \pi^-$, 
$\Sigma^+(1660)\to \pi^+ \Sigma \pi$  \cite{hemingway}.

\begin{table*}
\begin{center}
\caption{Table of results of the A-type fits, given in table \ref{table:a4pnewvalues}.
 The $\sigma_{\pi N}$ value enforced in the fits is given in the 
first row.
\label{table:a4newratios}}
\begin{tabular}{|c|r|r|r|r|}
\hline
$\sigma_{\pi N}$ & $20^*$    & $30^*$    &  $40^*$ \\
\hline
$\gamma$  & 2.36 & 2.36 & 2.37 \\
$R_c$   & 0.629 & 0.628&  0.628 \\
$R_n$   & 0.168 & 0.171&  0.173 \\
$\Delta E$ (eV) & 194 & 192 & 192 \\   
$\Gamma$  (eV)&  324 & 302& 270   \\
$\Delta E_D$ (eV) & 204 & 204 & 207 \\   
$\Gamma_D$  (eV)&  361 & 338& 305   \\
$a_{K^-p}$ (fm)& $-0.49+i\,0.44 $   & $-0.49+i\,0.41$  & $-0.50+i\,0.37$  \\
$a_0$      (fm)& $-1.07+i\,0.53 $   & $-1.04+i\,0.50$  & $-1.02+i\,0.45$ \\
$a_1$      (fm)& $0.44+i\,0.15 $   & $0.40+i\,0.15$  & $0.33+i\,0.14$    \\
$\delta_{\pi \Lambda}(\Xi)$  ($^\circ$) & 3.4& 4.5 & 5.7\\
$m_0$  (GeV) & 1.2 & 1.1& 1.0\\
$a_{0+}^+$  ($10^{-2}\cdot M_\pi^{-1}$)& $-2.0$  & $-2.2$ & $-2.2$\\
\hline
\end{tabular}
\end{center}
\end{table*}

 The number of data points included in each fit,  without 
the data for the energy shift and width of kaonic hydrogen, is 97. 
 Unless the opposite is stated, we also include  in the fits 
the DEAR measurement of the shift and width of the 
$1s$ kaonic hydrogen energy level \cite{DEAR},
\ba
 \Delta E&=&193\pm 37 (stat) \pm 6 (syst.) \hbox{ eV},\nn\\
\Gamma&=&249 \pm 111 (stat.)\pm 39 (syst.) \hbox{ eV}\, , 
\label{deardata}
\ea 
which is around a factor of two more precise than  the KEK 
previous measurement \cite{ito},
$\Delta E=323 \pm 63\pm 11$ eV and $\Gamma=407\pm 208 \pm 100$ eV.
To calculate the shift and width of the $1s$  
kaonic hydrogen state we use the results of \cite{akaki}
incorporating  isospin  breaking corrections up to and including 
 ${\cal O}(\alpha^4,(m_d-m_u)\alpha^3)$. 
   We further constrain our fits by computing at $\Opd$ in baryon SU(3) CHPT
several $\pi N$  observables with the values of the low energy
 constants  involved in the fit.  Unitarity corrections in the $\pi N$ sector
are  not as large as in the 
 $S=-1$ sector, e.g., there isn't anything like a 
$\Lambda(1405)$ resonance close to threshold,  and hence a calculation within 
pure SU(3) baryon CHPT  is more reliable here. Thus,  we calculate  at $\Opd$,    
  $a_{0+}^+$, the isospin-even pion-nucleon S-wave scattering length,
  $\sigma_{\pi N}$, the pion-nucleon $\sigma$ term, 
and $m_0$ from the value of the proton mass $m_p$. 
In this way we fix three of our free parameters.

\begin{table}[H]
 \begin{center}
\caption{B-type fits that do not agree with the DEAR data,
eq.(\ref{deardata}).
 The enforced $ \sigma_{\pi N}$ value in the fit is shown in the first line. 
 For precise definitions of the parameters $f$, $b_0$, $b_D$, $b_F$, $b_i$ and $a_i$ 
 see ref.\cite{oepja}. Three among the parameters
 $b_0$, $b_D$, $b_F$ and $b_i$ are fixed.
 \label{table:b4newvalues}}
\begin{tabular}{|l|l|r|r|r|r|}
\hline
Units& $\sigma_{\pi N}$ & $20^*$ & $30^*$ & $40^*$  & ${\Op}$ \\
     & MeV           &     &      &    &\\
\hline
MeV &         $f$      & $95.8$    & $113.2$  & $100.0$    & 93.9  \\
GeV$^{-1}$ & $b_0$     & $-0.201$  & $-0.159$  & $-0.487$  &  $0^*$  \\
GeV$^{-1}$ & $b_D$     & $-0.005$  & $-0.297$  & $0.127$   & $0^*$\\
GeV$^{-1}$ & $b_F$     & $-0.133$  & $-0.157$  & $-0.188$  & $0^*$   \\
GeV$^{-1}$ & $b_1$     & $+0.122$  & $+0.016$   & $+0.135$ & $0^*$\\
GeV$^{-1}$ & $b_2$     & $-0.080$  & $-0.151$   & $-0.037$ & $0^*$ \\
GeV$^{-1}$ & $b_3$     & $-0.533$  & $-0.281$  & $-0.494$  & $0^*$ \\
GeV$^{-1}$ & $b_4$     & $+0.028$  & $-0.291$  & $-0.173$  & $0^*$\\
	  & $a_1$      & $+4.037$  & $+4.188$  & $+2.930$  & $-2.958$\\
	  & $a_2$      & $-2.063$  & $-3.129$  & $-2.400$  & $-1.479$\\
	  & $a_5$      & $-1.131$  & $-1.214 $  & $-1.225$ & $-1.330$\\
	  & $a_7$      & $-3.488$  & $-3.000$  & $-2.795$  & $-1.805$\\
	  & $a_8$      & $-0.347$  & $+0.642 $  & $+2.906$ & $-0.655$\\
	  & $a_9$      & $-1.767$  & $-2.109$  & $-1.913$  & $-1.918$\\
 \hline
\end{tabular}
\end{center}
\end{table}

\begin{figure}
\caption[pilf]{\protect \small
 The  solid lines correspond to the $\sigma=40^*$ MeV fit,
the dashed lines  to the $30^*$ MeV fit, and the
 dash-dotted curves to the $20^*$ MeV one of table 
 \ref{table:a4pnewvalues}. The different lines can be barely distinguished. 
 For experimental references
 see ref.\cite{oepja}. 
\label{fig:a4pnew}}
\centerline{\epsfig{file=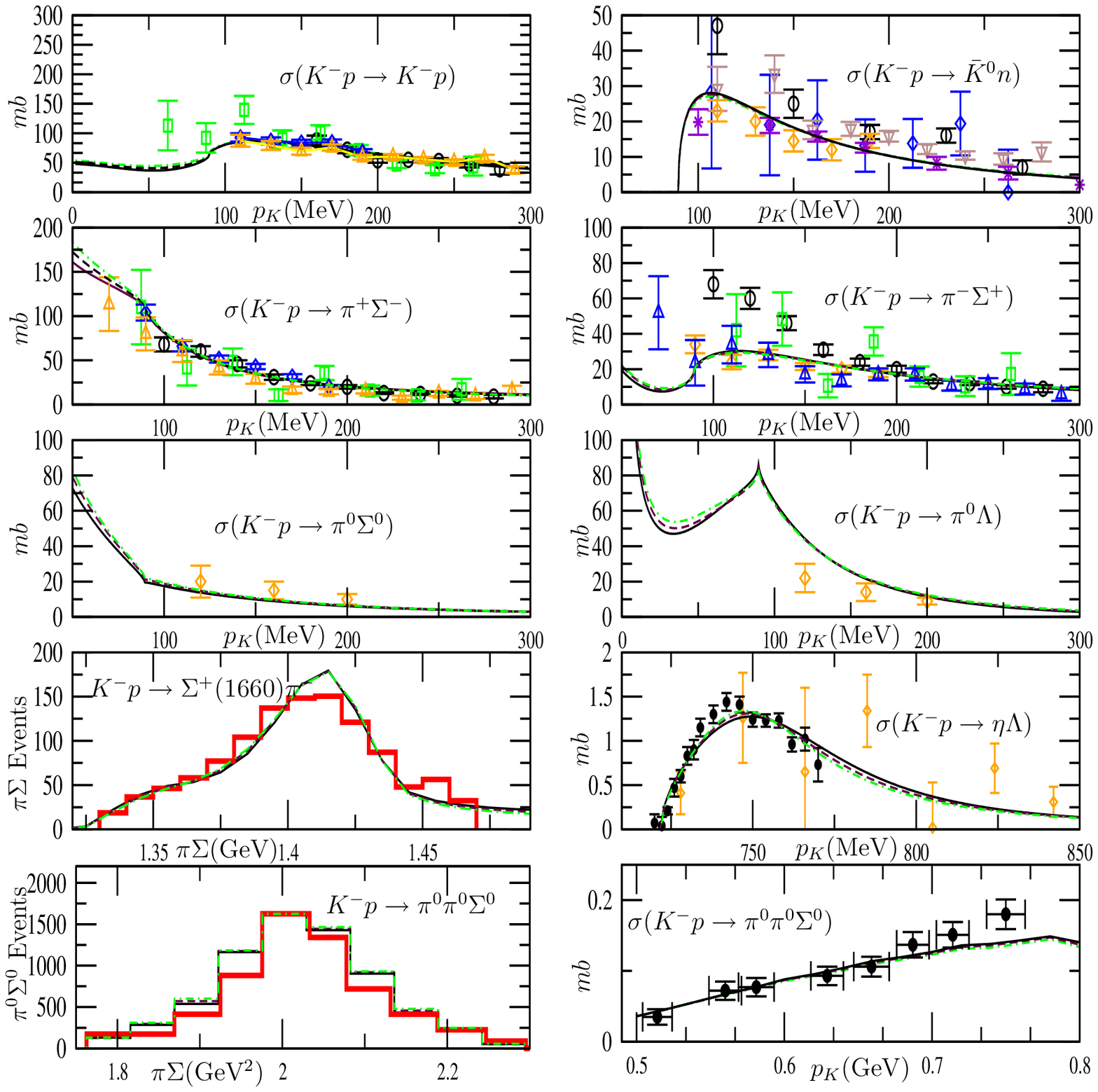,height=5.5in,width=3.8in,angle=0}}
\vspace{0.2cm}
\end{figure}

\begin{figure}
\caption[pilf]{\protect \small
 The solid lines correspond to the $\sigma_{\pi N}=40^*$ MeV fit, the dashes lines to the
  $30^*$ MeV fit, the dash-dotted curves to the $20^*$ MeV one and the dotted lines to the ${\Op}$ 
 fit of table \ref{table:b4newvalues}.
\label{fig:b4new}}
\centerline{\epsfig{file=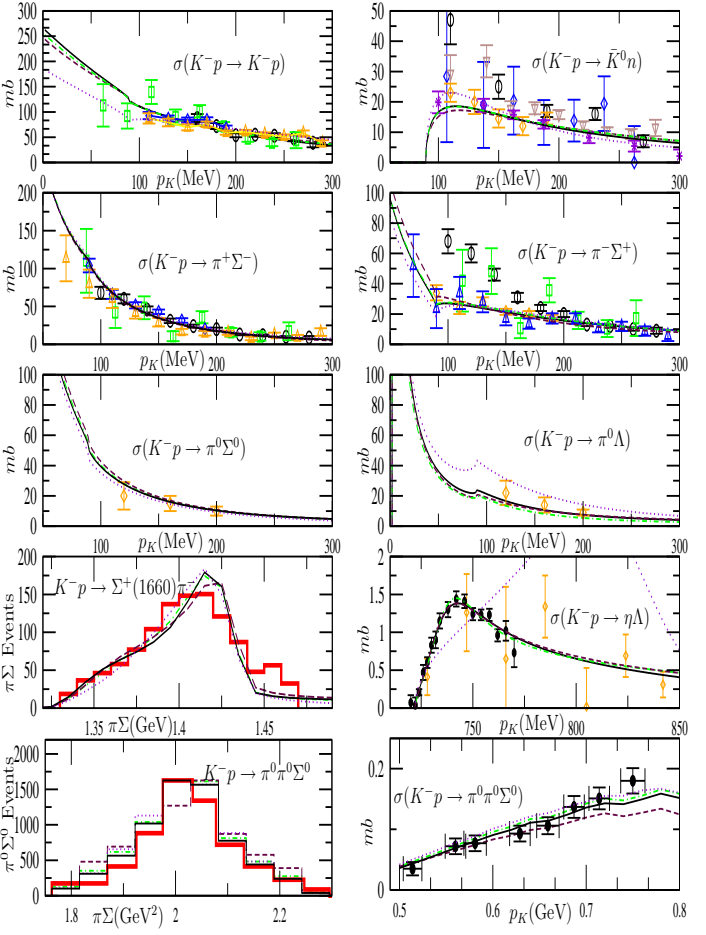,height=5.5in,width=3.8in,angle=0}}
\vspace{0.2cm}
\end{figure}

In addition ref.\cite{oepja} included further data from recent experiments of the 
Crystal Ball Collaboration \cite{nefkens,prakhov}. These data comprises the $K^-p\to 
\eta \Lambda$ cross section and $\Sigma \pi$ event distributions from the 
reaction $K^- p \to \pi^0 \pi^0 \Sigma^0$.  As noted in ref.\cite{oepja} these data cannot
be reproduced from the fits given in ref.\cite{opv} and then new fits were considered 
in the former reference that from the beginning included the data from 
\cite{nefkens,prakhov}. As in ref.\cite{opv} two type of fits were found. The so called 
A-type fits, table \ref{table:a4pnewvalues}, that together with scattering data 
also reproduce the DEAR measurement on kaonic hydrogen, and the B-type fits, table \ref{table:b4newvalues}, that reproduce the former 
but not the latter. In fig.\ref{fig:a4pnew} and table \ref{table:a4newratios} 
we show the reproduction of the data by the A-type fits and in fig.\ref{fig:b4new}
 and table \ref{table:b4newratios}  the same is done for the B-type fits. In the last column of 
table \ref{table:b4newvalues} we include the lowest order fit, only with $\ck_1$, also shown in 
fig.\ref{fig:b4new}.

\begin{table*}
\begin{center}
\caption{Table of results for the B-type fits, given in table \ref{table:b4newvalues}.
\label{table:b4newratios}}
\begin{tabular}{|c|r|r|r|r|r|}
\hline
$\sigma_{\pi N}$ & $20^*$    & $30^*$    &  $40^*$ & ${\Op}$\\
\hline
$\gamma$  & 2.34 & 2.35 & 2.34 & 2.32 \\
$R_c$   & 0.643 & 0.643&  0.644 & 0.637 \\
$R_n$   & 0.160 & 0.163&  0.176 & 0.193\\
$\Delta E$ (eV) & 436 & 409 & 450 & 348\\   
$\Gamma$  (eV)  & 614    & 681 & 591 & 611 \\
$\Delta E_D$ (eV) & 418 & 385 & 436 & 325\\   
$\Gamma_D$  (eV)  & 848 & 880 & 844 & 775 \\
$a_{K^-p}$ (fm)& $-1.01+i\,1.03 $   & $-0.93+i\,1.07$  & $-1.06+i\,1.02$  
& $-0.79+i\,0.94$ \\
$a_0$      (fm)& $-1.75+i\,1.15 $   & $-1.65+i\,1.30$  & $-1.79+i\,1.10$ 
& $-1.50+i\,1.00$\\
$a_1$      (fm)& $-0.13+i\,0.39 $   & $-0.14+i\,0.36$  & $-0.12+i\,0.46$ 
& $0.32+i\,0.46$   \\
$\delta_{\pi \Lambda}(\Xi)$  ($^\circ$) & $-1.4$ & 1.7 & $-1.2$ & $-1.4$\\
$m_0$  (GeV) & 0.8 & 0.6 & 0.7 & - \\
$a_{0+}^+$  ($10^{-2}\cdot M_\pi^{-1}$)& $-0.5$  & $-1.4$ & $+0.3$ & -\\
\hline
\end{tabular}
\end{center}
\end{table*}

It is worth stressing the good reproduction of the data accomplished by the A-type 
fits, including the most recent measurement of the kaon hydrogen lowest energy level and width. 
 One also observes a factor of 2 of difference between the $K^-p$ scattering lengths 
of the A- and B-type fits. So, if finally the DEAR measurement \cite{DEAR} is confirmed 
by the results of the DEAR/SIDDHARTA Collaboration \cite{sid}, it would give rise to an 
important step forward in the knowledge of kaon-nucleon interactions. This
difference in the scattering lengths 
 makes also that only the A-type fits have a pattern of isospin violation in 
the calculations of the shift and width of kaonic hydrogen of expected size \cite{akaki}. For
the B-type fits the isospin violations  turn out to be rather large, 30\%, while for the
A-type only 14\%. In addition, we also observe from tables \ref{table:a4newratios} and 
\ref{table:b4newratios} that the values of the scattering lengths are rather independent 
of the values given to  the sigma terms.

\section{Spectroscopy}

We show in tables \ref{table:a4pnewpoles0} and \ref{table:a4pnewpoles1} 
the I=0 and 1 poles, respectively, corresponding 
to the so-called fit I of ref.\cite{oepja}, 
given in the fifth column of table \ref{table:a4pnewvalues}.

\begin{table*}
\begin{center}
\caption{Fit I, I=0 Poles. The pole positions are given in MeV and the couplings in GeV. The symbol 
$|\g_i|_{\hbox{I}}$ means the coupling of the corresponding pole to the state with definite isospin 
I made up by the charged states of the $i_{th}$ channel. The couplings to
the I=1, 2 channels are always close to zero. 
\label{table:a4pnewpoles0}}
\begin{tabular}{|r|r|r|r|r|r|r|r|r|r|}
\hline
Re(Pole) & -Im(Pole) &Sheet & & & & & & & \\
 $|\g_{\pi\Lambda}|$ & $|\g_{\pi\Sigma}|_0$ & $|\g_{\pi\Sigma}|_1$ 
 & $|\g_{\pi\Sigma}|_2$ & $|\g_{\bar{K} N}|_0$ & $|\g_{\bar{K} N}|_1$ & $|\g_{\eta\Lambda}|$ 
 & $|\g_{\eta\Sigma}|$ & $|\g_{K\Xi}|_0$ & $|\g_{K\Xi}|_1$\\
\hline
1301 & 13 & 1RS & & & & & & &\\
0.03 & 1.12 & 0.02 & 0.01 & 5.83 & 0.05 & 0.41 & 0.04 & 2.11 & 0.03 \\
\hline
1309 & 13 & 2RS & & & & & & &\\
0.02 & 3.66 & 0.02 & 0.02 & 4.46 & 0.04 & 0.21& 0.04 & 3.05 & 0.03 \\ 
\hline
1414 & 23 & 2RS & & & & & & &\\
0.14 & 4.24 & 0.13 & 0.01 & 4.87 & 0.39 & 0.85& 0.20 & 9.35 & 0.11 \\ 
\hline
1388 & 17 & 3RS & & & & & & &\\
0.02 & 3.81 & 0.02 & 0.02 & 1.33 & 0.04 & 0.42& 0.04 & 9.55 & 0.04 \\ 
\hline
1676 & 10 & 3RS & & & & & & &\\
0.01 & 1.28 & 0.03 & 0.00 & 1.67 & 0.01 & 2.19 & 0.07 & 5.29 & 0.07 \\ 
\hline
1673 & 18 & 4RS & & & & & & &\\
0.01 & 1.26 & 0.02 & 0.00 & 1.82 & 0.01 & 2.13 & 0.06 & 5.32 & 0.06 \\ 
\hline
1825 & 49 & 5RS & & & & & & &\\
0.02 & 2.29 & 0.02 & 0.00 & 2.10 & 0.02 & 0.89 & 0.03 & 7.43 & 0.09 \\ 
\hline
\end{tabular}
\end{center}
\end{table*}

\begin{table*}
\begin{center}
\caption{Fit I, I=1 Poles. The pole positions are given in MeV and the couplings in GeV. The 
couplings to the I=0, 2 channels are always close to zero. The notation is like in table 
\ref{table:a4pnewpoles0}.
\label{table:a4pnewpoles1}}
\begin{tabular}{|r|r|r|r|r|r|r|r|r|r|}
\hline
Re(Pole) & -Im(Pole) &Sheet & & & & & & & \\
 $|\g_{\pi\Lambda}|$ & $|\g_{\pi\Sigma}|_0$ & $|\g_{\pi\Sigma}|_1$ 
 & $|\g_{\pi\Sigma}|_2$ & $|\g_{\bar{K} N}|_0$ & $|\g_{\bar{K} N}|_1$ & $|\g_{\eta\Lambda}|$ 
 & $|\g_{\eta\Sigma}|$ & $|\g_{K\Xi}|_0$ & $|\g_{K\Xi}|_1$\\
\hline
1425 & 6.5 & 2RS & & & & & & &\\
1.35 & 0.24 & 1.66 & 0.01 & 0.35 & 3.92 & 0.05 & 4.23 & 0.49 & 2.98 \\ 
\hline
1468 & 13 & 2RS & & & & & & &\\
2.80 & 0.16 & 5.96 & 0.02 & 0.23 & 8.74 & 0.04 & 10.66 & 0.19 & 2.48 \\ 
\hline
1433 & 3.7 & 3RS & & & & & & &\\
0.65 & 0.08 & 0.80 & 0.00 & 0.12 & 1.58 & 0.02 & 5.82 & 0.20 & 2.14 \\ 
\hline
1720 & 18 & 4RS & & & & & & &\\
1.82 & 0.02 & 1.21 & 0.00 & 0.02 & 0.95 & 0.02 & 6.78 & 0.05 & 5.31 \\ 
\hline
1769 & 96 & 6RS & & & & & & &\\
2.65 & 0.00 & 0.61 & 0.00 & 0.00 & 2.48 & 0.00 & 3.32 & 0.01 & 4.22 \\ 
\hline
1340 & 143 & 3-4RS  & & & & & & &\\
1.33 & 0.14 & 5.50 & 0.02 & 0.02 & 1.58 & 0.00 & 3.28 & 0.03 & 1.20\\
\hline
1395 & 311 & 3-4RS   & & & & & & &\\
2.08 & 0.01 & 1.49 & 0.01 & 0.00 & 1.24 & 0.00 & 7.63 & 0.01 & 3.97 \\
\hline
\end{tabular}
\end{center}
\end{table*}

By passing continuously from one Riemann sheet to the other some of the poles in 
the tables with the same isospin are
connected and represent the same resonance. One observes poles corresponding to the 
$\Lambda(1405)$, $ \Lambda(1670)$ and $\Lambda(1800)$ in good agreement with the mass and
width given to those resonances in the PDG \cite{pdg}. In addition, there is a lighter 
resonance around 1310 MeV, not quoted in the PDG, and this has to do with the 
so called two $\Lambda(1405)$ resonances, although for fit I it appears much lighter than in 
ref.\cite{teamL}. For I=1 one has the $\Sigma(1750)$  resonance. Fit I amplitudes also
show in I=1 a broad bump at around 1.6 GeV corresponding to the $\Sigma(1620)$. There are
also other poles around the $\bar{K}N$ threshold which mix up giving rise to a clear bump
structure from 1.4 to 1.45 GeV. These poles could be related to the $\Sigma(1480)$
\cite{cosy}. Finally, we also observe an I=2 pole for fit I at $1722 -i \,181$ MeV.

In tables \ref{table:b4newpoles0}
and \ref{table:b4newpoles1} the I=0, 1 poles positions for the fit II of ref.\cite{oepja}, 
given in the fifth column of table \ref{table:b4newratios}, are shown.
 There are also poles corresponding 
to the $\Lambda(1405)$, $\Lambda(1670)$ but not for the $\Lambda(1800)$. There is no
 $\Sigma(1750)$ resonance either and the bumps for the $\Sigma(1620)$ have disappeared
in several amplitudes.
\begin{table*}
\begin{center}
\caption{Fit II, I=0 Poles. The pole positions are given in MeV and the couplings in GeV. 
The notation is like in table \ref{table:a4pnewpoles0}.
\label{table:b4newpoles0}}
\begin{tabular}{|r|r|r|r|r|r|r|r|r|r|}
\hline
Re(Pole) & -Im(Pole) &Sheet & & & & & & & \\
 $|\g_{\pi\Lambda}|$ & $|\g_{\pi\Sigma}|_0$ & $|\g_{\pi\Sigma}|_1$ 
 & $|\g_{\pi\Sigma}|_2$ & $|\g_{\bar{K} N}|_0$ & $|\g_{\bar{K} N}|_1$ & $|\g_{\eta\Lambda}|$ 
 & $|\g_{\eta\Sigma}|$ & $|\g_{K\Xi}|_0$ & $|\g_{K\Xi}|_1$\\
\hline
1347 & 36 & 2RS & & & & & & &\\
0.02 & 6.48 & 0.12 & 0.02 & 2.60 & 0.10 & 1.42 & 0.01 & 0.32 & 0.07 \\
\hline
1427 & 18 & 2RS & & & & & & &\\
0.12 & 3.87 & 0.23 & 0.01 & 6.99 & 0.23 & 3.49 & 0.05 & 1.64 & 0.32 \\ 
\hline
1340 & 41 & 3RS & & & & & & &\\
0.07 & 5.92 & 0.08 & 0.01 & 0.62 & 0.08 & 2.33 & 0.01 & 0.75 & 0.04 \\ 
\hline
1667 & 8 & 4RS & & & & & & &\\
0.03 & 0.77 & 0.05 & 0.00 & 0.59 & 0.01 & 3.32& 0.02 & 12.17 & 0.08 \\ 
\hline
1667 & 8 & 5RS & & & & & & &\\
0.03 & 0.77 & 0.05 & 0.00 & 0.59 & 0.01 & 3.32& 0.03 & 12.17 & 0.06 \\ 
\hline
\end{tabular}
\end{center}
\end{table*}

\begin{table*}
\begin{center}
\caption{Fit II, I=1 Poles. The pole positions are given in MeV and the couplings in GeV. 
The notation is like in table \ref{table:a4pnewpoles0}.
\label{table:b4newpoles1}}\begin{tabular}{|r|r|r|r|r|r|r|r|r|r|}
\hline
Re(Pole) & -Im(Pole) &Sheet & & & & & & & \\
 $|\g_{\pi\Lambda}|$ & $|\g_{\pi\Sigma}|_0$ & $|\g_{\pi\Sigma}|_1$ 
 & $|\g_{\pi\Sigma}|_2$ & $|\g_{\bar{K} N}|_0$ & $|\g_{\bar{K} N}|_1$ & $|\g_{\eta\Lambda}|$ 
 & $|\g_{\eta\Sigma}|$ & $|\g_{K\Xi}|_0$ & $|\g_{K\Xi}|_1$\\
\hline
1399 & 41 & 2RS & & & & & & &\\
1.49 & 0.09 & 5.58 & 0.01 & 0.13 & 4.92 & 0.08 & 0.73 & 0.03 & 4.99 \\ 
\hline
1424 & 3.6 & 2RS & & & & & & &\\
0.54 & 0.14 & 1.58 & 0.00 & 0.20 & 1.17 & 0.10 & 0.61 & 0.04 & 3.76 \\ 
\hline
1311 & 122 & 3-4RS & & & & & & &\\
2.63 & 0.05 & 4.61 & 0.01 & 0.02 & 3.44 & 0.02 & 0.60 & 0.03 & 3.60 \\ 
\hline
1426 & 3 & 3RS & & & & & & &\\
0.56 & 0.04 & 1.18 & 0.00 & 0.07 & 0.77 & 0.04 & 0.61 & 0.02 & 3.74 \\ 
\hline
\end{tabular}
\end{center}
\end{table*}

In summary we have reviewed the works   of refs.\cite{opv,reply,oepja}. We have shown
two type of fits that agree with scattering experimental data but only one type agrees
with the DEAR measurement of kaonic hydrogen \cite{DEAR}. 
These latter fits are also those that offer a
remarkable agreement with spectroscopic information \cite{pdg}. Hence, taken into 
consideration the present experimental information, the so called fits A, table 
\ref{table:a4pnewvalues}, are preferred over the fits B, table \ref{table:b4newvalues}.

 This work has been supported in part by the MEC (Spain) and FEDER (EC) Grants
  Nos. FPA2003-09298-C02-01 (J.P.), FPA2004-03470 (J.A.O. and M.V.),  the 
  Fundaci\'on  S\'eneca grant Ref. 02975/PI/05  (J.A.O. and M.V.), the European Commission
(EC) RTN Network EURIDICE under Contract No. HPRN-CT2002-00311 and the HadronPhysics I3
Project (EC)  Contract No RII3-CT-2004-506078 (J.A.O.) and by  Junta de Andaluc\'{\i}a Grants Nos.
FQM-101 (J.P. and M.V.) and FQM-347 (J.P.).
% M.V. also acknowledges financial support from the 
% Fundaci\'on S\'eneca (Murcia).% and the Departamento de F\'{\i}sica  Te\'orica y del Cosmos,
% Universidad de Granada, for the warm hospitality.

% BibTeX users please use
% \bibliographystyle{}
% \bibliography{}

\begin{thebibliography}{}
\bibitem{DEAR} G. Beer  {\it et al.} [DEAR Collaboration], 
 Phys. Rev. Lett. {\bf 94}, (2005) 212302.\vs
\bibitem{sid} D. L. Sirghi and  F. Sirghi, (DEAR/SIDDHARTA Collaboration), 
 The physics of kaonic atoms at DAFNE, \\
 {\small http://www.lnf.infn.it/esperimenti/dear/DEAR\_RPR.pdf}.
% \\
%  {\sc C. Curceanu (DEAR/SIDDHARTA Collaboration)}, Precision
%measurements of kaonic atoms at DAFNE,\\
% {\small www.tp2.ruhr-uni-bochum.de/vortraege/workshops/trento05/Petrascu.pdf}
\bibitem{akaki} U.-G. Mei{\ss}ner, U. Raha and A. Rusetsky, 
 Eur. Phys. J. {\bf C35}, (2004) 349.\vs
\bibitem{deser}  S. Deser {\it et al.},
 Phys. Rev. {\bf 96},  (1954) 774; 
%%CITATION = PHRVA,96,774;%%
 T. L. Trueman,  Nucl. Phys. {\bf 26}, (1961) 57.\vs
%%CITATION = NUPHA,26,57;%%

\bibitem{dalitz} R. H. Dalitz and S. F. Tuan,  Phys. Rev. Lett. {\bf 2},  (1959) 425; 
%%CITATION  = PRLTA 2,425;%%
Ann. Phys. {\bf 8},  (1959) 100.
%%CITATION = APNYA 8,100;%

\bibitem{galileo}  A.~D.~Martin, N.~M.~Queen and G.~Violini,
 Nucl.\ Phys. {\bf B10}, (1969) 481;  P.~M.~Gensini, R.~Hurtado and G.~Violini,
   PiN Newslett. {\bf 13},  (1997) 291;  B.~Di Claudio, A.~M.~Rodriguez-Vargas and G.~Violini,
  Z.\ Phys. {\bf C3}, (1979) 75.
  %%CITATION = ZEPYA,C3,75;%%

\bibitem{martin} A.~D. Martin, Nucl. Phys. {\bf B179},  (1979) 33.\vs

%\cite{Buttgen:1985yz}
\bibitem{juelich}
  R.~Buttgen, K.~Holinde and J.~Speth,  Phys.  Lett. {\bf B163},  (1985) 305; 
 R.~Buettgen, K.~Holinde, A.~Mueller-Groeling, J.~Speth and P.~Wyborny,
  Nucl.\ Phys.  {\bf A506},  (1990) 586; 
  A.~Mueller-Groeling, K.~Holinde and J.~Speth,
  Nucl.\ Phys. {\bf A513},  (1990) 557.
  %%CITATION = NUPHA,A513,557;%%
  %%CITATION = NUPHA,A506,586;%%
  %%CITATION = PHLTA,B163,305;%%


\bibitem{hamaie}  T. Hamaie, M. Arima and K. Masutani, Nucl. 
Phys. {\bf A591}, (1995) 675 


\bibitem{landau} P.~J.~Fink, G.~He, R.~H.~Landau and J.~W.~Schnick,
  Phys.  Rev.  {\bf C41},  (1990) 2720.
  %%CITATION = PHRVA,C41,2720;%%



\bibitem{cloudy} E.~A.~Veit, B.~K.~Jennings, R.~C.~Barrett and A.~W.~Thomas,
  Phys.\ Lett. {\bf B 137},  (1984) 415.
  %%CITATION = PHLTA,B137,415;%%
  %%CITATION = NUCL-TH 9709023;%%
  %%CITATION = NUPHA,B10,481;%%

\bibitem{schat} J.~L.~Goity, C.~L.~Schat and N.~N.~Scoccola,
 Phys.  Rev. {\bf D66}, (2002) 114014;   
  Phys.\ Rev.\ Lett.  {\bf 88}, (2002) 102002.
   %%CITATION = HEP-PH 0111082;%%
  %%CITATION = HEP-PH 0209174;%%

\bibitem{kaisersiegel}N.~Kaiser, P.~B.~Siegel and W.~Weise,
 Nucl.\ Phys.  {\bf A594}, (1995) 325.
  %%CITATION = NUCL-TH 9505043;%%


\bibitem{npa} J.~A.~Oller and E.~Oset,
 Nucl.\ Phys.  {\bf A620}, (1997) 438; {\em (E)-ibid.}  {\bf A652}, (1999) 407.
  %%CITATION = HEP-PH 9702314;%%  
  

\bibitem{oset} E. Oset and A. Ramos,
  Nucl. Phys. {\bf A635},  (1998) 99.\vs


\bibitem{reportramos} J.~A.~Oller, E.~Oset and A.~Ramos,
 Prog.\ Part.\ Nucl.\ Phys.  {\bf 45}, (2000) 157.
  %%CITATION = HEP-PH 0002193;%%

\bibitem{ollerm}  J.~A. Oller and U.-G. Mei{\ss}ner,
 Phys. Lett. {\bf B500}, (2000) 263.\vs

\bibitem{teamL} D. Jido, J.~A. Oller, E. Oset, A. Ramos and U.-G. Mei{\ss}ner, 
  Nucl. Phys. {\bf A725}, (2003) 181.\vs


\bibitem{lutznieves}  C. Garcia-Recio, M.~F.~M. Lutz and J. Nieves,
 Phys.\ Lett.  {\bf B582}, (2004) 49.
  %%CITATION = NUCL-TH 0305100;%%


\bibitem{bura}
  B.~Borasoy, R.~Nissler and W.~Weise,
 Phys. Rev. Lett. {\bf 94},  (2005) 213401; Ibid. {\bf 96}, (2006) 199201;    
 Eur.  Phys.  J. {\bf A25}, (2005) 79; Phys. Rev. {\bf C74}, (2006) 055201. 
 arXiv:hep-ph/0606108.
  %%CITATION = HEP-PH 0512279;%%
  %%CITATION = HEP-PH 0410305;%%
  %%CITATION = HEP-P 0505239;%%
  %%CITATION = HEP-PH 0606108;%%


\bibitem{reply} J.~A.~Oller, J.~Prades and M.~Verbeni,
  Phys.\ Rev.\ Lett.\  {\bf 96}  (2006) 199202.
  %%CITATION = HEP-PH 0601109;%%


\bibitem{opv} J.~A. Oller, J. Prades and M. Verbeni, 
 Phys. Rev. Lett. {\bf 95}, (2005) 172502. 


\bibitem{oepja} J.~A.~Oller, Eur. Phys. J. {\bf A28}, (2006) 63.

\bibitem{nefkens} A.~Starostin {\it et al.}  [Crystal Ball Collaboration],
 Phys.\ Rev.\ {\bf C64}, (2001) 055205.
  %%CITATION = PHRVA,C64,055205;%%


\bibitem{prakhov}  S.~Prakhov {\it et al.}  [Crystal Ball Collaboration],
 Phys.\ Rev. {\bf C70}, (2004) 034605.
  %%CITATION = PHRVA,C70,034605;%%


\bibitem{magas} V.~K. Magas, E. Oset and A. Ramos,
 Phys.\ Rev.\ Lett.  {\bf 95}, (2005)  052301.
  %%CITATION = HEP-PH 0503043;%%

\bibitem{lag} J.~A.~Oller, M.~Verbeni and J.~Prades,  J. High Energy Phys {\bf 09}, (2006) 079;
 (E)-hep-ph/0701096;
M. Frink, U.-G. Mei{\ss}ner,  Eur. Phys. J. {\bf A29}, (2006) 255.
  %%CITATION = HEP-PH 0608204;%%

\bibitem{ito} M. Iwasaki {\it et al.},
\newblock Phys. Rev. Lett. {\bf 78}, (1997) 3067;
 Phys. Rev. {\bf C58},  (1998) 2366.\vs

\bibitem{26plb} W.~E. Humphrey and R.~R. Ross, Phys. Rev. {\bf 127},  (1962) 1305. 

\bibitem{27plb}J.K. Kim,  Phys. Rev. Lett. {\bf 14}, (1965) 29.

\bibitem{28plb}  M. Sakitt {\it et al.}, 
 Phys. Rev. {\bf 139}, (1965) B719.

\bibitem{31plb}  J. Ciborowski {\it et al.},  J. Phys. {\bf G8}, (1982) 13.

\bibitem{29plb}  W. Kittel, G. Otter and I. Wacek, Phys. Lett. {\bf 21}, (1966) 349. 

\bibitem{30plb} D.~Evans {\it et al.}, 
 J.\ Phys.  {\bf G9}, (1983) 885.
  %%CITATION = JPHGB,G9,885;%%

\bibitem{nowak} R.J. Nowak {\it et al.}, 
 Nucl. Phys. {\bf B139}, (1978) 61.
 
\bibitem{tovee} D. Tovee {\it et al.},  Nucl. Phys. {\bf B33}, (1971) 493.

\bibitem{hemingway}  R.J. Hemmingway, Nucl. Phys. {\bf B253}, (1985) 742.\vs

\bibitem{pdg} W.~M.~Yao {\it et al.}  [Particle Data Group],
  J.\ Phys.\  {\bf G33} (2006) 1.
  %%CITATION = JPHGB,G33,1;%%

\bibitem{cosy} I.~Zychor {\it et al.},
  Phys.\ Rev.\ Lett.  {\bf 96}, (2006) 012002.
  %%CITATION = NUCL-EX 0506014;%%

%Author, \textit{Book title} (Publisher, place year) page numbers
% etc
\end{thebibliography}

\end{document}